\documentclass[sigconf]{acmart}
\AtBeginDocument{%
  \providecommand\BibTeX{{%
    \normalfont B\kern-0.5em{\scshape i\kern-0.25em b}\kern-0.8em\TeX}}}

\setcopyright{acmcopyright}
\copyrightyear{2021}
\acmYear{2021}

\acmConference[UIST '21]{UIST 2021: ACM Symposium on User Interface Software and Technology}{October 10--13, 2021}{Virtual}
\acmBooktitle{UIST 2021: ACM Symposium on User Interface Software and Technology, October 10--13, 2021, Virtual}

\usepackage{xspace}
\newcommand{\mytitle}{MedKnowts\xspace}
\newcommand{\code}[1]{\texttt{#1}}
\usepackage{wrapfig}

\usepackage{adjustbox}

\begin{document}

\title{\mytitle: Unified Documentation and Information Retrieval for Electronic Health Records}

\settopmatter{authorsperrow=3}

\author{Luke Murray}
\email{lsmurray@mit.edu}
\affiliation{%
  \institution{MIT CSAIL}
  \streetaddress{32 Vassar Street}
  \city{Cambridge}
  \country{United States}
}

\author{Divya Gopinath}
\email{divyagopinath@alum.mit.edu}
\affiliation{%
  \institution{MIT CSAIL}
  \streetaddress{45 Carleton Street}
  \city{Cambridge}
  \country{United States}
}

\author{Monica Agrawal}
\email{magrawal@mit.edu}

\affiliation{%
  \institution{MIT CSAIL \& IMES}
  \streetaddress{45 Carleton Street}
  \city{Cambridge}
  \country{United States}
}

\author{Steven Horng}
\email{shorng@bidmc.harvard.edu}
\affiliation{%
  \institution{BIDMC}
  \streetaddress{1 Deaconess Road}
  \city{Boston}
    \country{United States}
}

\author{David Sontag}
\email{dsontag@csail.mit.edu}
\affiliation{%
  \institution{MIT CSAIL \& IMES}
  \streetaddress{45 Carleton Street}
  \city{Cambridge}
    \country{United States}
}

\author{David R. Karger}
\email{karger@mit.edu}
\affiliation{%
  \institution{MIT CSAIL}
  \streetaddress{45 Carleton Street}
  \city{Cambridge}
    \country{United States}
}





\renewcommand{\shortauthors}{Murray et al.}


\begin{abstract}
Clinical documentation can be transformed by Electronic Health Records, yet the documentation process is still a tedious, time-consuming, and error-prone process. Clinicians are faced with multi-faceted requirements and fragmented interfaces for information exploration and documentation. These challenges are only exacerbated in the Emergency Department---clinicians often see 35 patients in one shift, during which they have to synthesize an often previously unknown patient’s medical records in order to reach a tailored diagnosis and treatment plan. To better support this information synthesis, clinical documentation tools must enable rapid contextual access to the patient’s medical record. MedKnowts is an integrated note-taking editor and information retrieval system which unifies the documentation and search process and provides concise synthesized concept-oriented slices of the patient’s medical record. MedKnowts automatically captures structured data while still allowing users the flexibility of natural language. MedKnowts leverages this structure to enable easier parsing of long notes, auto-populated text, and proactive information retrieval, easing the documentation burden.

\end{abstract}




\maketitle

\def\DK#1{\textcolor{blue}{DK: #1}}

\section{Introduction}

Electronic Health Records (EHRs) have been adopted in the hope that they would improve quality of care, save time, support collaboration and data sharing, and prevent clinical errors \cite{Davidson2004, batesImprovingSafetyInformation2003, friedbergFactorsAffectingPhysician2014,schnipperSmartFormsElectronic2008}. However current EHR platforms have largely failed to achieve these goals. Studies of EHR adoption have shown both positive and negative effects \cite{mamykinaClinicalDocumentationComposition2012,kuhnClinicalDocumentation21st2015}, but clinicians now spend more time navigating EHRs than physically communicating with patients and EHR usage is a leading cause of physician burnout and stress \cite{menachemiBenefitsDrawbacksElectronic2011, sieglerPrioritizingPaperworkPatient2015, moyMeasurementClinicalDocumentation2021}. 

Despite being laborious to create, well-written clinical documentation is invaluable. At their best, cogent clinical narratives can help clinicians understand a patient's case \cite{patelImpactComputerbasedPatient2000, mamykinaClinicalDocumentationComposition2012}, function as a powerful communication method between clinicians \cite{coieraWhenConversationBetter2000}, and serve as learning tools to improve future care practice \cite{batesImprovingSafetyInformation2003}. But EHRs rarely achieve this and arguably interfere with it. The issue lies in the fragmentation among views in the EHR for the two processes underlying the clinical workflow: (i) information retrieval and data exploration over a patient's history and (ii) information entry. 
Because structured and unstructured data can be hard to reconcile, EHRs often store and display information in separate pages or windows, and physicians have to synthesize the patient narrative by navigating across a variety of sources \cite{ruleValidatingFreetextOrder2015, ahmedEffectTwoDifferent2011}. This creates increased cognitive burden to discover unstructured information, and studies have shown that clinicians spend more time reading past notes than doing any other activity in the EHR \cite{calvittiTemporalAnalysisPhysicians2012}. Further, the fragmented interfaces hinder comprehensibility and necessitate frequent task-switching \cite{zhengInterfacedrivenAnalysisUser2009, mamykinaClinicalDocumentationComposition2012, coieraWhenConversationBetter2000}. To avoid this context switching, clinicians have developed coping mechanisms such as copying from previous notes or using autofill techniques for naive pre-population of text \cite{haasCliniciansPerceptionsUsability2005,makamUseSatisfactionKey2013,ruleClinicalDocumentationEndUser2020}. Unfortunately, indiscriminate use of these auxiliary functions causes documentation to become bloated, making it difficult for clinicians to parse important clinical information, and potentially even propagating errors \cite{tuttyComplexCaseEHRs2019,ruleClinicalDocumentationEndUser2020,hartzbandRecordAvoidingPitfalls2008, weisCopyPasteCloned2014}.

\subsection*{Our Contribution}

\begin{figure*}
  \begin{center}
    \includegraphics[width=0.98\textwidth]{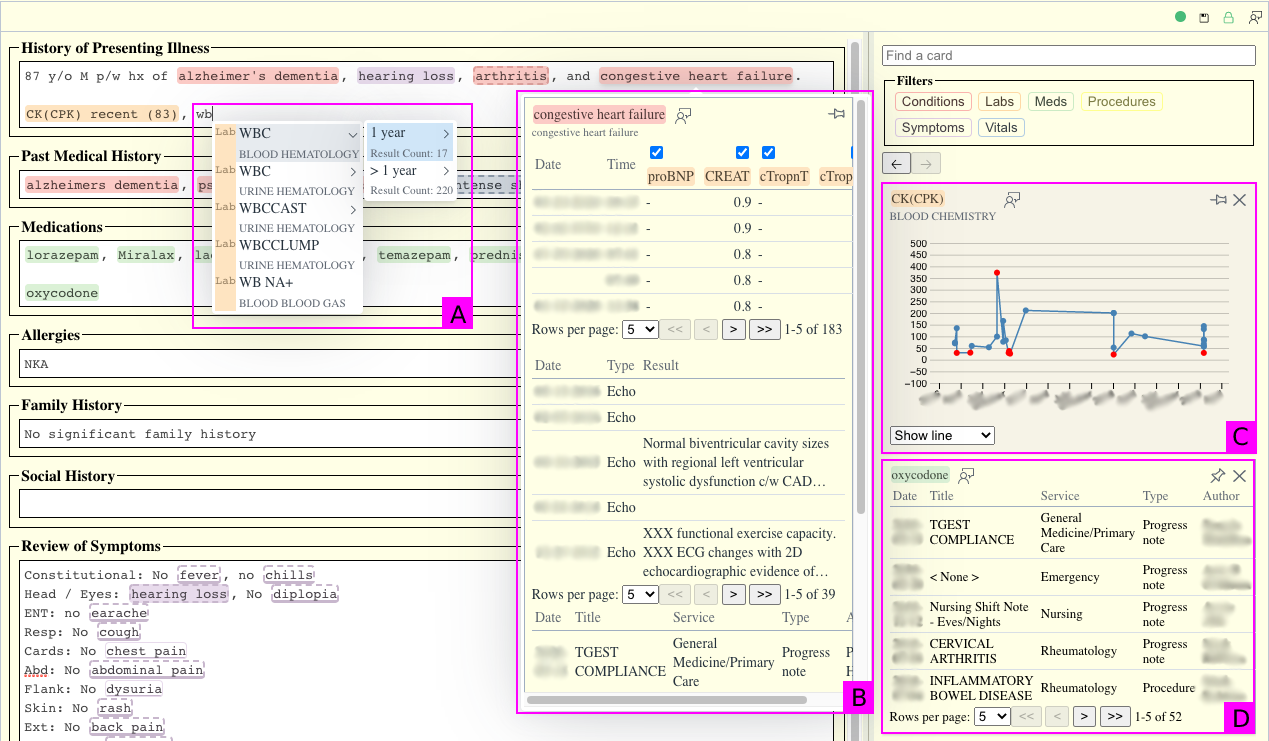}
  \end{center}
 \Description{Screenshot of the presented system, showcasing multiple features. On the left, a free text form with multiple sections is augmented by a dropdown menu for inserting lab values, a cardiac card that is hovered over. On the right, the sidebar contains a search bar, a graph of lab values in one card, and relevant notes to oxycodone use in a card below that.}
  \caption{The \mytitle interface containing sections of the clinical note on the left, and an integrated sidebar on the right. The user is typing \code{WBC} and triggering autocomplete (A). Detail text in the autocomplete is used to differentiate clinical terms and provide additional context such as result counts. The card for the most recent identified term, \code{CK}, is displayed in the preview pane (C) with values displayed as a line chart, and abnormal values highlighted in red. The preview pane history can be navigated using the backwards and forwards buttons at the top of (C). Below the preview pane the doctor has pinned a card for \code{oxycodone} (D), which displays note snippets relevant to \code{oxycodone}. A search bar at the top of the sidebar can be used as an alternative method to add cards to the preview pane. A transcluded card for cardiac conditions (B) shows labs, cardiology reports, and note snippets relevant to congestive heart failure in a single interface.}
  \label{fig:interface_annotated}
\end{figure*}


In this work, we propose a novel documentation system for EHRs, \mytitle, which passively assists clinicians by seamlessly integrating an editor for clinical documentation with a \emph{concept-oriented view}\cite{zengKnowledgebasedConceptorientedView2001} of the patient's medical history. 
\mytitle  provides contextual autocomplete (Fig. \ref{fig:interface_annotated}A) for clinical terms (e.g. conditions, symptoms), saving precious documentation time.
The autocomplete works without a trigger character---so it does not disrupt the prior documentation workflow---and displays options for structured data entry (e.g. lab values) as the user types, removing the need to memorize content importing phrases. When autocomplete is not used, we employ keyword matching, which we call \textit{post recognitions}, to automatically identify clinical terms as the clinician types.
Both auto-completed and post-recognized terms are transformed into structured interactive elements which we call \emph{chips}. We leverage this structure for live semantic highlighting that enables easier parsing of long notes and for automatic population of repetitive text fields, easing documentation burden. Therefore, \mytitle retains some of the benefits of structured data entry, while still allowing users the flexibility of natural language to describe the subtleties of complex patient narratives.

Further, we use the structured data to automatically surface \emph{information cards} in an attached preview pane (Fig. \ref{fig:interface_annotated}C) as the doctor types. Proactively displayed cards provide concise summaries of relevant medical history, reducing the context-switching required to synthesize a note.
Each card is a concept-oriented view \cite{zengKnowledgebasedConceptorientedView2001} such that information is grouped by underlying concept (e.g. the labs, medications, and notes related to a condition) rather than by data modality (all medications at once). Concept-oriented views have been shown to help physicians work faster and make fewer errors \cite{semanikImpactProblemorientedView2021}.
In addition to the automatically surfaced cards, chips embedded in the note and in cards serve as links to related cards, providing direct access to the relevant medical history from the note context and other cards. Cards can be surfaced in-line by hovering on a chip (Fig. \ref{fig:interface_annotated}B) or in the preview pane by clicking on a chip. This provides an additional avenue for contextual information retrieval without dividing attention between views. Finally, cards can be pinned to an attached sidebar (Fig. \ref{fig:interface_annotated}D), which persists the card to a view shared by the clinical care team, allowing for easier bookmarking, collaboration, and data sharing without directly copying to contribute to note bloat.

We present the following contributions to enhance the EHR note taking experience:
\begin{itemize}
\item We provide passive and automatic methods to insert and disambiguate clinical terms as the note is written and transform them into chips---interactive, structured elements which provide information scent about recognized vocabulary, semantic highlighting, access to inline documentation, and contextual information retrieval. We therefore retain benefits of structured data entry without sacrificing the flexibility or ease of natural language.
\item We augment the EHR note-taking interface with a shared sidebar to which clinicians can pin cards. Each card presents a concept-oriented view for a particular clinical term. The sidebar provides clinicians with a shared and persistent space, integrated with the documentation interface, where they can add and remove cards. It thereby situates, beside the semi-structured note, a collaborative, customizable, and context-specific view of structured data in a patient's medical record.
\item We proactively display a preview card of the most recently identified concept which updates as the user types. The preview card provides a consistent passive display of detailed information immediately relevant to the clinician's current decision making context, reducing the need for the physician to manually forage for information.
\item We present findings from a year long iterative prototyping and design process and a one month evaluation with four medical notetakers.
\end{itemize}
We implemented these designs in a prototype system which we deployed live among scribes in an Emergency Department (ED) at a Level I trauma center and tertiary, academic, adults-only, teaching hospital. Our system was designed over the course of a year, in collaboration with an emergency physician with over a decade of experience designing and deploying EHRs, and with ongoing feedback from stakeholders including scribes, medical students, and physicians. In practice, scribes found \mytitle easy to use with a quick learning curve and and indicated that they would use it frequently. Further, they found the features of \mytitle well-integrated, saving them time over their previous workflows both for documentation and information retrieval.

\section{Related Work}

\subsection{Information Capture}
Early EHRs were expected to transform clinical care by transitioning medical records from manually-organized and paper-based to automatic and digitized \cite{athertonDevelopmentElectronicHealth2011}. Many early EHRs were built around forms and structured data entry in order to capture structured records, but few modern EHR systems retain these designs \cite{rosenbloomDataClinicalNotes2011}. Structured data entry is far more cumbersome and time consuming to input than unstructured text \cite{tangeMedicalNarrativesElectronic1997,rosenbloomDataClinicalNotes2011,mcdonaldBarriersElectronicMedical1997}. Clinicians prefer recording information with unstructured narrative \cite{ginnekenPhysicianFlexibleNarrative1996, rectorFoundationsElectronicMedical1991} because of the increased expressivity of free-text \cite{wilcoxUsingNaturalLanguage2002b, johnsonElectronicHealthRecord2008}.
However even clinicians who want the flexibility and efficiency of free-text when documenting information prefer structure and standards when revisiting old notes to parse the patient's medical history \cite{rosenbloomDataClinicalNotes2011,johnsonElectronicHealthRecord2008}. 
\mytitle lets clinicians seamlessly access and capture structured patient information and clinical terms while writing free text narrative. \mytitle additionally synthesizes the existing patient medical record into concept-oriented cards which provide the clinician with a standardized and structured view of data extracted from a pre-existing EHR system.

\subsubsection{Automatic Term Recognition}

Most clinical recognition systems are designed for post-processing rather than real-time analysis \cite{savovaMayoClinicalText2010, HealthcareNaturalLanguage2021}. They extract structured information from unstructured narrative and free text after it has been authored~\cite{johnsonElectronicHealthRecord2008,rosenbloomDataClinicalNotes2011}.
Systems such as \textit{Doccurate} \cite{sultanumDoccurateCurationBasedApproach2019} have been designed to validate, augment, and visualize post-processed labels but few systems close the loop and enable clinicians to take advantage of identified structure in the medical note during the process of documentation \cite{rosenbloomDataClinicalNotes2011}.

Of the few proposed EHR paradigms that do implement real-time entity recognition during notewriting, they either fail to map to standard clinical ontologies \cite{billmanMedicalSensemakingEntity2007}, neglect to use this structured data capture to support clinical decision-making \cite{johnsonElectronicHealthRecord2008}, or do not provide concept disambiguation (Fig. \ref{fig:post_recognitions}) which is crucial given the overloading of medical terminology and limited accuracy of post-hoc clinical concept recognition \cite{savovaMayoClinicalText2010, billmanMedicalSensemakingEntity2007, johnsonElectronicHealthRecord2008, HealthcareNaturalLanguage2021}. \textit{Active Notes} \cite{wilcoxMinimizingElectronicHealth2011} inspires the design of several features in \mytitle such as tagging clinical concepts and displaying related information in an attached sidebar. However \textit{Active Notes} requires users to manually initiate data queries and tag concepts with a hot key, and does not visually distinguish clinical vocabulary until it is tagged, making it hard for clinicians to learn the recognized vocabulary. In contrast \mytitle is designed to passively and automatically assist users without active participation. \mytitle provides live semantic syntax highlighting for clinical terms indicating concept type, negations, and potential ambiguities; and automatically transforms autocompleted and post-recognized clinical terms into interactive chips which can be used to resolve ambiguities, and view relevant patient information inline as a tooltip or persisted in an integrated sidebar.

\subsubsection{Structured Data Capture}
Many modern EHRs support multiple modalities for inserting structured data into the note \cite{makamUseSatisfactionKey2013}. Some tools support carry-forward techniques where data is copied or paraphrased from previous notes \cite{haasCliniciansPerceptionsUsability2005a}; others let clinicians insert structured values into the note by clicking in the patient's history or typing special characters to trigger macros \cite{wilcoxPhysiciandrivenManagementPatient2010, ruleClinicalDocumentationEndUser2020}. Still others require the user to specify the template structure using a complex interface of forms \cite{banNovelUseDiscrete2017}.

\mytitle differs from previous carry-forward techniques \cite{ruleClinicalDocumentationEndUser2020} by autofilling using information captured earlier in the note, rather than limiting autofill to information that appears in the patient's prior medical record. This is particularly pertinent to documentation in an ED environment, since clinicians often have to repeat information within the same note in order to meet regulatory and billing requirements, and previous notes may not be applicable to the current visit, let alone exist. 

\mytitle supports structured data capture for clinical terms (conditions, symptoms, medications), lab results, and vital signs with a machine learning-driven autocomplete interface based on Gopinath et al. \cite{gopinathFastStructuredClinical2020}. 
The autocomplete interface displays completions of clinical terms as the user types, which provides information scent for the available clinical vocabulary.
Structured data capture is a common feature in EHRs often referred to as \textit{dotphrases} because the data is conventionally inserted with a phrase that starts with a period (e.g. \code{.meds}) \cite{ruleClinicalDocumentationEndUser2020}.
\mytitle differs from previous systems because the structured data insertions do not require a trigger character or memory of content-importing phrases.
Trigger characters were unpopular in our deployments, since they require foresight to enter and knowledge of valid phrases.

Additionally, structured data templates, documented in Rule et al. \cite{ruleClinicalDocumentationEndUser2020}, work well in medical specialties such as ophthalmology, where many standard structured measurements are taken before the patient sees the doctor. However, in clinical settings such as the ED, the vast majority of structured data entry opportunities are contextually dependent on information needs arising after the clinician begins documentation. Thus, our more fluid workflow for structured data insertion within narrative text is an important extension to Rule et al.’s structured templates.

\subsection{Information Fragmentation in EHRs}

Studies of EHR usage have shown that separation of documentation interfaces from patient data cause clinicians to frequently task switch, creating cognitive overload and increasing the likelihood of clinical errors \cite{mamykinaClinicalDocumentationComposition2012,parkerImprovingClinicalCommunication2000, ahmedEffectTwoDifferent2011}.
Some previous EHR systems attempt to resolve this by presenting the entire medical record next to the documentation interface in complex interface of tabs, lists, and tables \cite{johnsonElectronicHealthRecord2008,farriImpactPrototypeVisualization2012}.
These interfaces are hard to parse, require manual navigation, and leave the complex work of synthesizing data from across the medical record to the clinician \cite{mamykinaClinicalDocumentationComposition2012}.
Other EHR systems, such as the one in use at the hospital in which we deployed, provide dashboards summarizing high value information  next to documentation \cite{ahmedEffectTwoDifferent2011, pickeringImplementationClinicianDesigned2015}.
In an ED these summary displays rarely include all the information clinicians need to access throughout the course of a visit. 
Still other research systems allow users to interactively filter a view of the patient's medical record to display data relevant to a particular concept\cite{wilcoxPhysiciandrivenManagementPatient2010,schnipperSmartFormsElectronic2008,hirschHARVESTLongitudinalPatient2015}.
These systems allow users to filter by one concept at a time and do not persist the data for later reference.

\mytitle lets clinicians access a curated subset of the medical record, displayed as a collection of concept-oriented cards.
Each card provides a succinct display of high value information curated for a single clinical concept.
The card relevant to the most recently recognized term is automatically displayed next to the note in a preview pane, providing a passive stream of relevant information to the clinician.
Previous work has shown that  clinicians are much less likely to perform manual actions to see information \cite{yangUnremarkableAIFitting2019}.
Cards can also be manually pinned to the sidebar where they can be seen by all users working on the note.
Pinned cards act as a persistent and shared collection of data which is particularly pertinent to a given patient's context.

\subsection{Problem-oriented Medical Records}

In the early 1970s Weed proposed the notion of problem-oriented medical records \cite{weedMedicalRecordsThat1968}. In the problem-oriented medical record, all information is organized around patient problems. Problem-oriented medical records were designed to reflect the way the physician thinks \cite{tangeHowApproachStructuring1996}, but did not survive. A major reason for their failure is that they require physicians to enter and maintain data organized around problems---often requiring multiple steps to input a single piece of data, while competing chronologically-oriented medical records offered unstructured text entry which was lightweight and fast in comparison \cite{tangeMedicalNarrativesElectronic1997}. 

Problem-oriented medical records (POMR), problem-oriented views, and concept-oriented views are very similar but have slight distinctions. Problem-oriented medical records refers to original idea proposed by Weed \cite{weedMedicalRecordsThat1968} to organize medical records around a problem list.
Problem-oriented views (POV), introduced by Buchanan \cite{buchananAcceleratingBenefitsProblem2017}, dynamically generate problem-oriented displays of information from a traditionally organized medical record. POVs do not require the user to input information organized around problems.
POVs place the the burden of organizing information around problems on the computer not the user.
Concept-oriented views (COV) introduced by Doré \cite{doreObjectOrientedComputerbased1995}, are an extension of POVs to all concepts not just problems.

\subsection{Information Foraging Theory}
Information foraging theory draws parallels between how humans hunt for information and how animals hunt for food---in particular, it identifies that users rarely find information in a completely linear process. Instead, useful information often appears in patches for which the user must forage, using clues in the user interface referred to as information scent \cite{pirolliInformationForagingInformation1995}.

Previous research into information foraging theory in EHRs highlights that the value of clinical information is not intrinsic but rather dynamic and task-specific  \cite{gibsonForagingInformationEHR2017}. Information that is relevant and important for one patient during one visit may not be relevant or important for another patient or in another clinical context. \mytitle presents a consistent stream of context-specific cards in a preview pane. Each card is analogous to an information patch, and the user can quickly determine if the card is worth foraging in and exploiting by reading the card title or scanning the card content which is consistent across cards. If a card is useful, the user can exploit the information patch by persistently pinning the card to their sidebar.

\mytitle encodes information scent within the documentation interface by providing semantic syntax highlighting for clinical terms in the form of chips. Terms are colored based on their concept type, whether or not they are negated, and whether or not they collide with other terms---these clue the user into how we have inserted structure into the note and what downstream information benefits to expect. Small visual indicators next to clinical terms, and detail text in the UI provide additional information scent and inform clinicians about whether a card is likely to contain information from the patient's medical record. Users can easily navigate between information patches. Users can navigate to cards by clicking or hovering on clinical terms embedded within both cards and the note taking interface, or by searching for a clinical term in the sidebar.

\section{Environment}
\subsection{Clinical Workflow}
While \mytitle was designed for use in a particular hospital's ED, here we describe a high-level clinical workflow that is generally common across EDs.
During a typical day in a hospital ED, clinicians may evaluate, treat, and document up to 35 patients. 
The note is used for various purposes: as a tool for communication and collaboration between present and future clinicians; as a document of the evidence-based decision making process the clinician utilizes to construct a care plan; and as a record for legal and reimbursement purposes \cite{Davidson2004, CentersforMedicareandMedicaidServices2020}. Before the clinician evaluates a patient, a triage nurse first prioritizes a patient, taking vital signs, assigning a chief complaint, and writing a brief triage note. The clinician then evaluates the patient and reviews the patient's prior medical record. As in almost all healthcare settings, time is limited and must be balanced between bedside care and reviewing the patient medical record. The main sections expected in the final documentation then closely mirror the underlying clinical workflow after triage \cite{CentersforMedicareandMedicaidServices2020}:

\noindent \textit{History of Present Illness (HPI).} The HPI serves as a chronological narrative of the patient's reason for the visit, including the presence, onset, severity, and duration of symptoms. Additionally, it involves surfacing medical history that may be relevant for contextualizing the patient's condition. Unlike in specialties that provide longitudinal care, emergency visits are episodic and unscheduled; emergency physicians are often meeting a patient for the first time, forcing them to quickly synthesize a patient's medical background from various sources, including past medical records.
\\
 \noindent \textit{Review of Symptoms (ROS).} The ROS contains an inventory of symptoms, documented per body system (e.g. cardiovascular, gastrointestinal). Information from the HPI is often repeated here.\\
\noindent  \textit{Medical Decision Making (MDM).} MDM is the complex process by which the clinician reaches a diagnosis and treatment plan. Within the MDM section, physicians need to enumerate the differential diagnosis, consider risks associated with various diagnostic and treatment options, and settle on the labs, tests, medications, and scans that must be conducted as part of the workflow. 

The sections above provide a comprehensive view of the patient's visit by corresponding to the systematic and thorough process behind patient evaluation and management. However, there is often overlapping information in the sections above due to billing requirements \cite{1997DocumentationGuidelines}, e.g. the ROS may include symptoms that were already mentioned in the HPI, the MDM often contains elements of the past medical history, leading to complaints of excessive, often repetitive data entry \cite{Kroth2018}.

\subsection{A Variety of Documentation Processes}
The documentation process described here is based on observations at the ED in which \mytitle was deployed. Some aspects of this process, such as the use of scribes, may not generalize to other EDs.
There is marked inter- and intra-provider variation in the processes to reach the final documentation based on individual clinician preferences, resources, and schedules.
Some clinicians write the majority of notes after their shift, jotting details during to jog memory later.
In addition to the final note, there exists an additional \textit{Clinician Comment} box which can be used for such intermediate thoughts, and is often additionally used as scratch space between members of the care team (e.g. an attending physician, a resident, a medical student) that are not part of the medical record. Others choose to write the majority during the shift, only revisiting the notes to make small edits and submit their notes to the official record. 

On another dimension, alternatives to keyboard text entry include (i) the use of voice dictation software and (ii) the employment of a scribe.
Scribes shadow the clinician, recording what they observe during patient encounters as well as discussions with other clinicians, and drafts notes for each of the patients that the clinician is seeing.
These notes are then handed over to the clinician, who will edit and augment to prepare the note for official recording in the patient's medical record. 
Since a lot of information communicated during the visit is irrelevant to the patient's care, the scribe acts as a filter that determines, documents, and relays clinically-relevant information. Experienced scribes may even search and synthesize the patient's past medical records themselves.
Because scribes were already writing notes at the ED \mytitle was deployed in, they were the target subjects for our study.
However in other hospitals where clinicians act as their own scribes, the clinicians would be the target users.
Voice dictation software can be used as an alternative text entry method when scribes are not available. But voice dictation does not fulfill other roles the scribe performs in the clinical workflow. 
In this study, due to incompatibilities in the deployed commercial dictation software, we specifically focused on scribe-physician workflow.
However, we note that interaction with dictation software is an infrastructural challenge and not a fundamental obstacle to using our system. 

\subsection{Study Environment}
The study described in this work was performed within a single Level I trauma center and tertiary, academic, adults-only, teaching hospital which provides care for 55,000 patients per year. The existing deployed web-based EHR was custom developed at the institution, but uses a commercially available documentation module. The study was approved by our institutional review board with a waiver of informed consent.

\mytitle was developed through prototypal deployments over the course of a year, during which a clinician and the clinician's scribes used the tool as their predominant note system. We report on lessons learned from the iterative prototyping process, as well as usage data collected from a one month long deployment at the end of the year.

\section{Design and Implementation}

The overall goal of our system is to reduce the effort clinicians must invest in retrieving information from the EHR, synthesizing that information into knowledge, and recording it into patient notes.  We do so via a combination of interacting features:

\begin{enumerate}
    \item We use \emph{autocomplete} as well as \emph{post recognition} to recognize meaningful concepts from a large, standard medical ontology.  Autocomplete can save users keystrokes.  More importantly, these standard concepts provide an indication of the problem the clinician is addressing for the current patient and are inserted as structured \emph{chips}.
    \item We use the recognized concepts to pre-populate other portions of the note that require duplication of that information, relieving clinicians and scribes of that burden.
    \item We introduce a \emph{preview pane} and persistent \emph{sidebar} for delivery of standardized \emph{cards} of contextual information relevant to recognized concepts.  When a concept is recognized, the relevant card is automatically introduced in the preview pane, proactively providing clinicians with information they are likely to need to address the problem whose description they are currently typing in their note.  Cards also group and organize this information to help clinicians gain insight about long-term trends and associations.
    Clinicians can additionally pin cards to the sidebar to create a persistent shared collection of information pertinent to the patient context.
    \item We provide all these affordances with a passive and automatic design, which does not require active participation from the user.
\end{enumerate}

In this way, we can simultaneously decrease documentation burden on physicians and use the captured clinical terms to aid physicians in information retrieval while typing a note. We elaborate on these features below.

\subsection{Autocomplete}

\begin{figure}
  \begin{center}
    \includegraphics[width=0.98\columnwidth]{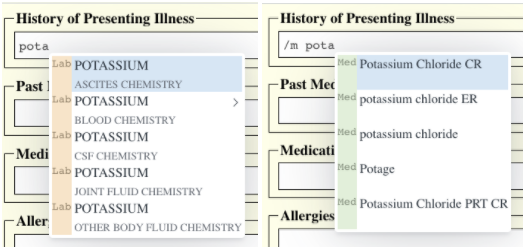}
  \end{center}
  \Description{On the left, a dropdown showing only potassium-related labs after "pota" is typed. On the right, a dropdown showing different potassium-related labs after "/m pota" is typed.}
  \caption{Autocomplete in the same context without filters and with filters. The "/m" command is used to limit the clinical terms displayed to medications.}
  \label{fig:autocomplete_filter}
\end{figure}

The backbone of the structured data capture within \mytitle is a contextual autocomplete mechanism. We hypothesized autocomplete would enable structured clinical data capture without disrupting the existing documentation workflow and potentially even decrease keystroke burden on clinicians. Autocompleted terms could then be used to facilitate information retrieval and clinical decision support, offering longer term benefits.

We bootstrapped our autocomplete with a subset of clinical terms pulled from the SNOMED and UMLS medical ontologies \cite{Bodenreider2004, NIH-NLM2015}. The ontologies contain abbreviations and synonyms for each term, allowing users to employ the language they are most comfortable with. 

In our initial prototypes we use a single character trigger \code{/} to start the autocomplete, similar to \textit{dotphrases} commonly found in EHRs \cite{ruleClinicalDocumentationEndUser2020}.
When triggered, the autocomplete displayed a dropdown filtered to terms whose prefix matched the characters following the initial trigger. The clinician and scribes disliked the trigger because it required foresight that they were entering structured data or typing a recognizable concept and a priori knowledge of the set of recognized concepts. When no suitable term existed, users had to manually delete the trigger character.

Therefore, our next iteration, outlined in Gopinath et al. \cite{gopinathFastStructuredClinical2020}, replaced the character trigger with a collection of rule-based triggers based on particular phrases, word boundaries, and punctuation.
As an example within this paradigm, the phrase ``presents with" is likely followed by a symptom, so the algorithm will show the autocomplete dropdown with symptoms listed first.
User feedback indicated that rule based ranking is insufficient---the autocomplete often failed to display desired terms; and the boundary and punctuation triggers cause autocomplete to appear, unnecessarily, distracting the user. 

To improve on the rule-based approach, we replaced the rules with a one-dimensional convolutional neural network model that predict when to trigger, and what type of clinical concept to prioritize, since a learned model can encode nuanced syntactical relationships. It significantly outperforms the rule-based triggering approach described in \cite{gopinathFastStructuredClinical2020}, achieving a precision of 43\% versus 7\%. Precision is defined as the fraction of times the user wanted to type a clinical concept when the autocomplete was triggered. In addition, after optimization, inference of this model requires an average autocomplete latency of about 18 milliseconds, which is close to the screen refresh rate and therefore perceived as instantaneous to the user.

While the model based approach works well, users indicated a desire to manually override the model---either forcing autocomplete to trigger or specifying the clinical concept to rank first.
In these cases, we resort to slash filters: \code{/labs} or \code{/l} can be used to trigger an autocomplete context which is limited to labs.
An empty slash forces autocomplete to trigger with the default ranking.
An example of why filtering is useful can be seen in Figure \ref{fig:autocomplete_filter}. These filter shortcuts give users the fine-grained ability to easily insert structured information at any place in the note. 

\subsection{Post Recognitions}

During prototyping users disliked that \mytitle only identified clinical concepts entered with autocomplete.
Unrecognized terms could appear because the user opted not to use autocomplete or because the user pasted text into the note.
This issue was particularly noticeable when we used recognized terms to pre-populate later sections of the note.
Some scribes would spend time re-entering unrecognized terms using autocomplete because they perceived the unrecognized term to be an error or wanted to generate the correct text later in the note.
To resolve these issues, we implemented a version of the Aho-Corasick algorithm to automatically identify clinical terms from the text that has already been typed \cite{Aho1975}. We dub this tagging mechanism \emph{post recognition}. 

\subsection{Semantic Highlighting and Concept Disambiguation}

As clinical jargon is notoriously overloaded, it is often the case that the same string can describe multiple terms \cite{smith_overloaded_jargon}. For example, \textit{Pt} can refer to a patient, physical therapy, or prothrombin time. While clinicians generally have the domain expertise to disambiguate between similar terms, jargon can create confusion for patients, medical trainees, and clinicians of a different specialty \cite{Wang2012}. Therefore, \mytitle needs to be able to correctly disambiguate each written term to its underlying clinical concept in the ontology in order for users to reap the benefits of contextual information retrieval features that our system offers.

\begin{figure}
  \begin{center}
    \includegraphics[width=0.7\columnwidth]{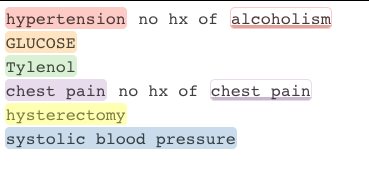}
  \end{center}
  \Description{Sample text from a medical note displaying six clinical terms, each highlighted in a different color.}
  \caption{Autocomplete inserts terms as highlighted immutable chips. They can be deleted, moved around and copied like other text, but they cannot be modified.}
  \label{fig:autocomplete_chips}
\end{figure}

\mytitle uses live syntax highlighting to provide visual information scent about terms the system recognizes. 
\mytitle supports six concept types: conditions, labs, medications, symptoms, procedures, and vital signs.
When the user accepts an auto completion, the system inserts a \textit{chip}---a highlighted block of text that can be copied, moved around, or deleted like other text.
Each chip is highlighted with a color associated with its concept type; an example from each of the six concept types \mytitle supports can be seen in Figure \ref{fig:autocomplete_chips}.

\begin{figure}
  \begin{center}
    \includegraphics[width=0.98\columnwidth]{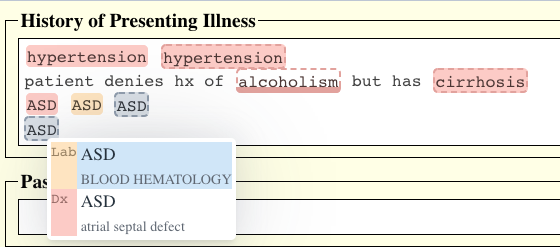}
  \end{center}
  \Description{A group of clinical terms, each highlighted with a dotted border and an in-line dropdown to choose between several underlying structured concept options.}
  \caption{Post recognitions are automatically recognized clinical terms. They are rendered with a dotted border and can be disambiguated through a popup menu on click. Negated post recognitions are rendered with an underline.}
  \label{fig:post_recognitions}
\end{figure}

Post recognized phrases are also replaced with chips.
However, while auto-completed phrases map to unique ontology items specified by the user's selection, post-recognized phrases can be ambiguous.
In the case that a post recognition requires disambiguation, the user can click on the chip to select from the relevant set of candidate terms.
Post recognitions are differentiated from autocomplete chips with a dotted border. When possible, the border also indicates the concept type with color:
if multiple clinical terms match a post recognition but each clinical term is from the same concept type, the color for that concept type is applied to the entire post recognition.
If clinical terms from multiple concept types match the post recognition then we display the recognition with a grey background.
An example can be seen in figure \ref{fig:post_recognitions}.

Clinicians often reference clinical terms to indicate the absence of something, for example "no fever".
In our initial prototyping we used double click to toggle chips between "positive" and "negated".
When negated the chip is highlighted with an underline, and the text is transformed---for example "fever" becomes "no fever".
Additionally, we provided autocompletions for each clinical term prefixed with "no" so that users could insert negated chips with autocomplete, but clinicians found this method of indicating negations brittle and disliked that lists of negated terms such as "no A, B, or C" had to be written as "no A, no B, no C" to comply with \mytitle' simple negation implementation.
To resolve these issues we implemented a modified version of negex \cite{chapmanSimpleAlgorithmIdentifying2001} to automatically identify and highlight negated chips based on the surrounding text.

In the autocomplete dropdown ambiguity can arise when a string refers to multiple terms. For example, \code{potassium} refers to multiple labs measured with various fluids, so we display this disambiguating information as detail text in the dropdown, as seen in Figure \ref{fig:autocomplete_filter}.

\subsection{Context-specific information retrieval}
To further aid clinicians, we automatically retrieve and display context-specific information from a patient's medical record. As an example, when a medication, procedure, or condition appears in the autocomplete dropdown, we use detail text---"in patient medical record" to indicate whether it previously appeared in the patient's medical record. We provide similar information scent next to chips with a small grey circle indicator.

\begin{figure}
  \begin{center}
    \includegraphics[width=0.98\columnwidth]{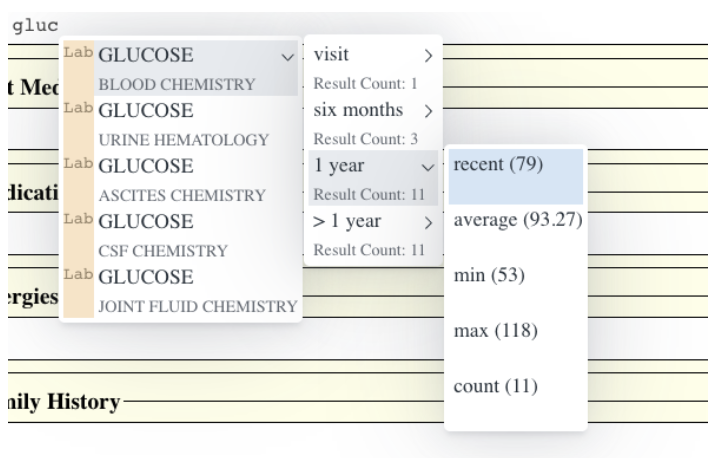}
  \end{center}
  \Description{Screenshot of the system showing a multi level tree menu expanded to fill in glucose values from the past year, after "gluc" has been typed.}
  \caption{An example of context-specific information retrieval. Autocomplete insertion of lab results using a tree based menu with support for aggregation at multiple time frames and specific values}
  \label{fig:autocomplete_tree}
\end{figure}

This structured retrieval and display is particularly handy for documenting labs---after receiving requests to automatically insert quantitative lab results using autocomplete, we implemented a tree-based lab selection menu, displayed in figure \ref{fig:autocomplete_tree}. This hierarchical menu can be used to insert structured data associated with an autocomplete term. The user can select the name of the lab, a time frame based aggregate, or individual statistics within a time frame. The time frame aggregate is inserted as a string \code{LAB\_NAME (MIN\_VALUE - MAX\_VALUE) AVG\_VALUE} and individual statistics are inserted as a string \code{LAB\_NAME STAT\_NAME STAT\_VALUE}. We also added the ability to insert vitals (pulse, heart rate, etc.) using the same methods, completing our set of clinical concept types.

\subsection{Default Text}
\begin{figure}
  \begin{center}
    \includegraphics[width=0.98\columnwidth]{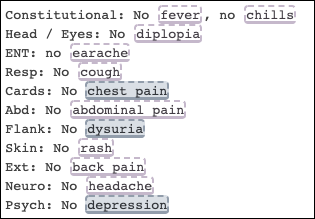}
  \end{center}
  \Description[]{Text from the review of systems section with ten lines one for each system followed by symptoms associated with that system.}
  \caption{An example of the review of systems section}
  \label{fig:ros_example}
\end{figure}

Medical notes are often pre-filled with boilerplate default text, but this text is often overwritten because it does not incorporate enough patient-specific context. \mytitle further reduces data entry by taking advantage of structured data capture and using it to fill in later sections of the note. To this end, we created templates for each of the sections of the notes based on clinician input. When the user clicks on a blank note section the section is autopopulated with the template text, which is constructed using a mix of structured information parsed from the patient's medical record as well as clinical terms previously captured in the note. As an example, the Review of Systems (ROS) section (Fig. \ref{fig:ros_example}) is a boilerplate list of ten systems, and for each system the clinician has to describe the presence or lack of symptoms related to that system. \mytitle automatically generates this ROS text for the clinician from text entered in previous sections---when a symptom is documented in the note, it is added to the appropriate line of the ROS template.

The addition of pre-populated text brought additional feedback from clinicians.
Clinical terms are often associated with clarifying modifiers and specifiers and it is important to retain these modifiers and specifiers when copying forward clinical terms.
For example "left lower abdominal pain" is more informative in diagnosing a condition than simply "abdominal pain".
Expanding the ontology to include all possible combinations of modifiers for each term is not feasible.
Instead we use a simple greedy algorithm to attach modifiers as prefixes to clinical concepts.
This algorithm could be replaced with more advanced NLP methods but we chose to use this lightweight approximation to satisfy run-time requirements.
The use of algorithms to detect negated and modified terms helps retain the nuance and meaning of the original text when copied across sections.

\subsection{Concept-Oriented Views}\label{subsec:concept-oriented}
Although there are multiple documentation systems in use at the hospital for writing ED notes, none of them are integrated with tools to view the patient's prior medical history. While some documentation systems provide limited views of a patient's information for the current hospital visit (e.g. recent labs or imaging), this does not help a clinician with reviewing and synthesizing the medical history. In order to access additional data, clinicians must still navigate through multiple different pages.

Some clinicians place two browser windows side by side and access data in one window and their note in another, others flip between pages and use their short-term memory to synthesize information. Both paradigms are error-prone---clinicians evaluate multiple patients in a shift and can easily navigate to the wrong patient's data or mis-remember details of patients with similar presentations. In addition, when interesting data such as a relevant note or lab trend is found by a clinician, there is no way to bookmark it for later use. All the computers in the hospital implement session timeouts to prevent the inadvertent sharing of patient information, so clinicians copy potentially relevant data into their note to preserve it and the surrounding context is lost.

\mytitle reduces the need for clinicians to hunt for and retrieve data from multiple sources by proactively fetching relevant data and surfacing it just-in-time. To achieve this, we introduce the notion of a \emph{card} for each clinical term in our ontology. Cards unify diverse information fragments related to the term in a single, templated, format.  
Each card has a header with the common name for the clinical term and synonyms for the clinical term from our ontology:
\begin{itemize}
    \item Condition cards (e.g. \code{diabetes})---display relevant medications from the patient's medical record, relevant vital signs, related procedures, and relevant snippets from notes in the patient's medical record.
    \item Labs and Vitals cards (e.g. \code{creatinine}, \code{blood pressure})---display a box and whisker chart of lab values.
    \item Procedures and Medications cards (e.g. \code{hysterectomy}, \code{metformin})---contain a list of relevant note snippets from the patient's medical history.
\end{itemize}
Note snippets are surfaced if they contained a mention of the term or a closely linked term and are ordered chronologically. The set of closely linked terms was algorithmically mined and a sample was validated by a clinician. Based on feedback, we excluded symptoms from our set of cards, as clinicians rarely needed medical history to contextualize symptoms.

\subsection{Surfacing Cards}

In our early prototypes we displayed cards in an attached sidebar when clinicians clicked on an associated chip within the note or another card.
However this created a two step process to see any card---first type the term with autocomplete and then select the term to see the card.
To reduce friction we automatically added a card to the sidebar for any term inserted with autocomplete.
However autocomplete is a poor signal for whether a card is useful is in the long term.
Cards added to the sidebar are displayed in a scrolling vertical stack.
Cards can be removed, but left alone, they persist next to the note for the duration of the note authoring process, and useful cards can be pushed out of view as more cards are added. 
Some clinicians found this method of adding cards to the sidebar unintutive or confusing, and other clinicians felt like they were seeing too much irrelevant information.
Additionally this method fails to surface post recognitions.

We eventually streamlined our approach to surfacing sidebar cards to a two-step process. Any time a term is recognized before the user's selection, we display the card for that term in a preview pane at the top of the sidebar. The preview pane displays one card at a time, and the card is not shared between users. Clinicians can pin a card displayed in the preview pane to move it to the sidebar. The cards pinned in the sidebar are persistent and are shared between multiple users. In this way, the sidebar becomes a collaborative record of the fragments from the patient's medical history that clinicians identify as being particularly important or relevant.

Cards are surfaced in the preview pane in one of three ways: first, they are automatically displayed when an autocompleted or post-recognized term appears before the user's selection; second, they are manually surfaced by users clicking on a chip within the note or another card; third, they are manually surfaced via a search bar at the top of the sidebar. Post recognitions with naming collisions (e.g. \code{pt}) must be disambiguated by the user before the associated card is surfaced. 

\subsection{Hand Designed vs Automatically Generated Cards}

Ideally, we could create individually designed and physician curated cards for all possible clinical concepts.
But we do not have the resources to take that approach.
Instead, during initial prototyping we created a meta-cards for each clinical concept (labs, conditions) which act as templates for all clinical terms within that clinical concept.
We describe the contents of the meta-card for each clinical concept in Section \ref{subsec:concept-oriented}.

Automatically generated cards help solve a cold-start problem, as we hypothesized that users would be unlikely to adopt the system if the majority of clinical terms were associated with empty cards.
But cards generated for a large number of clinical terms are slow to iterate on.
For example, clinicians asked for certain labs to be added to cardiac cards.
This type of change, if abstracted to all conditions, requires the development of a dataset to relate labs and conditions.
While possible, finding or creating this type of dataset takes time.
Conversely, adding lab values manually to cardiac cards is light weight and easy to validate with users.
In the long term hand designed features could be replaced with generic models or datasets but in the short term we can iterate faster by taking a manual approach.

\subsection{Card Design}

\begin{figure}
  \begin{center}
    \includegraphics[width=0.98\columnwidth]{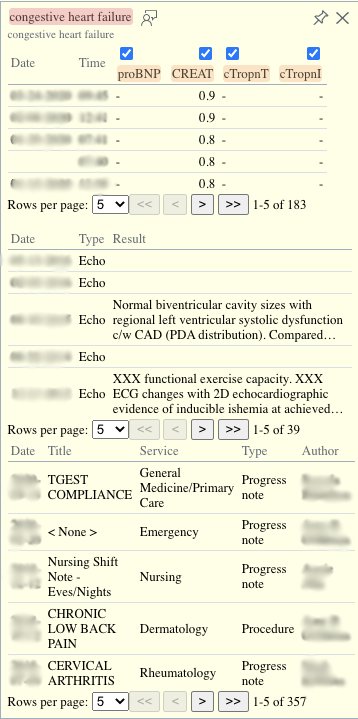}
  \end{center}
  \Description{An example card for cardiac conditions containing a table with 4 lab measurements taken across time, a table with recent echocardiographs alongside a report snippets, and a table of notes that mention the condition.}
  \caption{An example card surfaced for \textit{Congestive Heart Failure}, which contains pertinent lab values, links to recent echocardiography reports, and recent notes that mention the condition.}
  \label{fig:cardiac_card}
\end{figure}

Throughout the prototyping process clinicians consistently displayed strong preferences about the content of cards.
We hypothesized that showing synonyms within cards would familiarize users with our ontology of terms. But clinicians found the inclusion of synonyms condescending because they \textit{already knew that}.
We received a similar response when we listed names of labs related to a condition on condition cards.
However, clinicians reacted positively when we listed the names of labs along with their values since this is proactively fetching relevant information.
Clinicians want to see information relevant to their decision making and other information is seen as noisy or unnecessary.

In addition, clinicians want information presented in the immediate format that they require; as an example, if the most recent lab value is the only useful piece of information, that is the only lab value that should be displayed. Conversely, some lab values can only be properly interpreted in the context of other lab values. For example, interpreting an elevated troponin values requires both prior troponin values and prior creatinine values. In that case both lab values must be displayed. We provided feedback forms on cards and accumulated various requests for data to be displayed on particular cards. However implementing granular changes for generic classes of cards is difficult..

To address this, in our second iteration of cards we chose to specifically focus on two types of cards: lab cards, and cards related to cardiac conditions. In the long run we expect that a set of a few thousand cards targeting individual clinical terms as well as general classes of clinical information (such as cardiac function) could support clinician's needs. While it is beyond our capacity to create an exhaustive set of cards, we can learn about and demonstrate the value of cards by creating a few for common terms. If proven valuable, other cards could be created by a small engineering team with clinical guidance, or even by clinicians themselves if given suitable authoring tools. 

We worked in collaboration with three physicians to design a card which presents information relevant to cardiac conditions. Our cardiac card includes labs and snippets from cardiac tests (EKG and Echocardiogram) and other free-text notes. An example of the cardiac card can be seen in Figure \ref{fig:cardiac_card}.

We augmented our lab card template to support multiple views of lab results. A table view can be used to see individual result values. When applicable, contextual labs that are useful for understanding the primary lab are added as columns to the table display. A zoomable line chart displays lab values over time and a box and whisker plot is used to display aggregate lab values over various time frames.
Additionally we provided support for contextual lab results in the table view.
For example, Kidney failure, which is measured by an elevated creatinine, leads to a build up of potassium, causing elevated potassium levels, a life-threatening condition that must be treated immediately. Whenever an abnormal potassium level is encountered, the next piece of information that is needed is what the kidney function is. We proactively provide this information by displaying creatinine levels directly on the potassium lab card.

\subsection{Inline Display of Cards}

Early on we realized it would be useful to access cards from within the note itself. We added the ability to hover on a chip to see a preview of the card.

\section{Evaluation}

As described previously, \mytitle was deployed in two major iterations---one year of iterative prototyping and a one month evaluation. For approximately 7 months the prototypal deployments were used as the primary documentation tool by 1 physician (who is also a co-author) and 4 scribes across 1185 patients; the evaluation lasted 1 month and was used by the same physician and 4 scribes (2 scribes had participated in the prototypal deployments) across 234 patients. Our prototypal deployment ended after the hospital stopped using scribes in the wake of COVID-19; the second deployment began soon after scribes returned to the hospital. We could not do a comparative study against the baseline documentation system due to legal limitations disallowing modifying the commercial note taking tool in use at the hospital, but we describe our evaluation below.

\vspace{5mm}

\begin{table}
\begin{tabular}{l|l|l|l}
User & Patients & Shifts & Pins \\ \hline
P    & 150      & 12                     & 58          \\
S1   & 69       & 3                       & 1           \\
S2   & 50       & 4                      & 0           \\
S3   & 43       & 2                      & 1           \\
S4   & 33       & 3                       & 15          \\ \hline
Totals   &          &                     & 75         
\end{tabular}
\Description{}
\captionof{table}{General usage data.  Totals for Patients and Shifts are left out because the scribes worked with the physician on the same patient/shift. Pins reflect the number of cards each user pinned to the sidebar.}
\label{tab:usage-data}
\end{table}
\hspace{.04\textwidth}
\begin{table}
\centering
\begin{tabular}{l|l|l}
User & Autocomplete & Post Recognition \\ \hline
P    & 71           & 7                               \\
S1   & 35           & 0                               \\
S2   & 27           & 0                               \\
S3   & 6            & 0                               \\
S4   & 4            & 0                               \\ \hline
Totals & 143          & 7                              
\end{tabular}
\captionof{table}{How users inserted chips. Autocomplete indicates insertion via autocomplete and post recognition is disambiguation of post recognitions.}
\label{tab:insert-chips}
\end{table}
\hspace{.04\textwidth}
\begin{table}
\centering
\begin{tabular}{l|l|l|l|l}
User & Search & Autocomplete & Post Recognition & Note Snippet \\ \hline
P    & 127    & 30         & 21                     & 3                  \\
S1   & 4      & 4          & 25                     & 0                  \\
S2   & 0      & 0          & 9                      & 0                  \\
S3   & 0      & 1          & 7                      & 0                  \\
S4   & 43     & 0          & 12                     & 21                 \\ \hline
Totals & 174    & 35         & 74                     & 24                
\end{tabular}
\captionof{table}{How users added cards to the sidebar. Autocomplete means clicking on an autocomplete chip, post recognition means clicking on a post recognition, search means performing a search in the sidebar, and note snippet means clicking on a note snippet in a card. Clicking on a note snippet displays a card containing the full note text with the snippet highlighted.}
\label{tab:viewed-cards}
\end{table}

\begin{figure*}[hbt!]
  \begin{center}
    \includegraphics[width=0.98\textwidth]{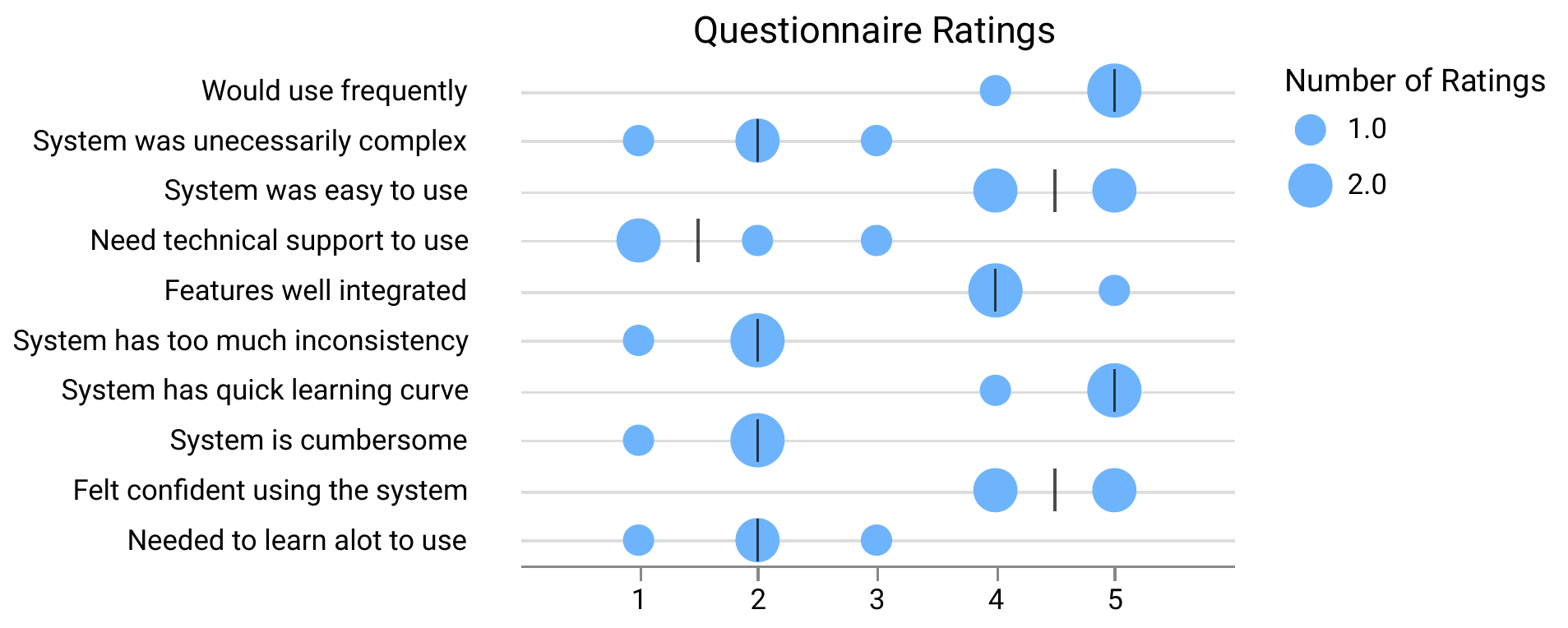}
  \end{center}
  \caption{Scribes' System Usability Scale scores with medians displayed as black bars for each question}
  \label{fig:scribe_likert_d1}
  \Description{Usability scores indicating that the system was generally sufficiently easy to use with a quick learning curve, and features were considered well-integrated and useful. 5 indicates strong agreement and 1 indicates strong disagreement.}
\end{figure*}

Prior to using the tool live in the ED, the scribes were introduced to the tool in thirty minute training sessions.
In each training session, one co-author showed the scribe how the tool worked and explained its available features.
After working in the ED the same co-author followed up with the scribes to get their feedback.
At the time of the follow up the scribes each had used the tool for an average of 3 shifts (min 2, max 4) and completed an average of 46.5 notes (min 33, max 69). In the study follow-up, scribes filled out a system usability scale (SUS) \cite{jordanUsabilityEvaluationIndustry1996} as seen in Figure \ref{fig:scribe_likert_d1}, and answered questions from a script.

The final SUS scores were \code{[77.5, 77.5, 85, 95]} (avg. 83.75), the physician did not fill out a SUS scale. A score in the high 70s to upper 80s is considered to be good while a score above 90 is excellent \cite{bangorEmpiricalEvaluationSystem2008}. These responses indicate that scribes found the tool relatively intuitive and useful enough to use frequently.

Feature usage, documented in Table \ref{tab:usage-data}, \ref{tab:insert-chips}, and \ref{tab:viewed-cards} as well as qualitative interviews yielded several takeaways.
In both the tables and the rest of the evaluation we refer to the users as Physician (P) or Scribe 1-4 (S1-S4).

Most scribes described that autocomplete sped up their workflows but adoption of autocomplete changed based on scribe experience.
S1, the least experienced scribe in the evaluation, noted that they liked autocomplete because they no longer needed to conduct internet searches to find correct spellings and obtain an understanding of the underlying concept space.
S4, the most experienced scribe, found autocomplete less useful due to familiarity with terms, but still found utility for longer terms.

As users acclimated to the tool's functionality their usage changed.
For example, S1 used autocomplete 8 times in their second shift and 27 times in their third shift.
The increase in usage in the third shift was due to the use of autocomplete to insert lab values.
It was unclear if the scribe had discovered this functionality on the third shift or had become familiar enough with the tool to adopt more advanced features.

Some of the differences in feature usage across scribes may be attributed to discoverability.
For example S2 found card transclusion to be very helpful, especially for getting more familiar with unknown terms while S4 did not realize that they could hover on chips to see cards inline.
S3 stated that card transclusion was helpful to quickly hover and get a sense of how central the concept is to the patient's history.

The lack of disambiguations for post recognitions may be due to the fact that post recognized chips both behave and look very similar to chips inserted with autocomplete.
For example, if an ambiguous term is highlighted correctly and copied appropriately in default text, the scribe may not have any need to disambiguate it.
Both S3 and S4 were happy that the system recognized terms but were unaware that post-recognized terms could be disambiguated.

Scribes appreciated the colored highlighting of embedded chips in the notes.
They found that it allowed them to quickly scan what had occurred so far.
For example, they could quickly skim through symptoms to orient themselves, and it was helpful that negated symptom mentions were visualized differently.
One scribe mentioned that they could use the colors as an automatic visual aid to determine what components had been completed in the Medical Decision Making section, and what was left to be documented.
This quick skim approach wasn't necessary for certain concept types (e.g. medications), but some scribes still found it useful for organization. 

Scribes universally appreciated the default text that was auto-populated due to the structured data capture from autocomplete and post recognitions. This was most appreciated in Physical Exam and Review of Systems, despite imperfections in the default text. One scribe (S3) said it ``made them much more efficient" allowing them to ``get through charts faster." Another noted that the checkbox-based systems employed in the hospital's commercial EHR made it really easy to skip and miss an item, indicating the new system felt less error-prone due to its data entry.

At a high level, scribe experience correlated to the amount of synthesis of a patient's past history that was conducted, as advanced scribes had accrued more of the requisite clinical knowledge and reasoning and could handle documentation and synthesis simultaneously.
S4 was the only scribe to examine past notes to try and find relevant information to share with the clinician.
S4 liked the note snippets stating that "it saves me a lot of time compared to reviewing all of the patient's prior notes to simply be able to click...and have notes show up."
In contrast S2 did not pay much attention to cards or read note snippets because they were focusing on documenting what the clinician was saying.

More experienced scribes were more likely to pin cards to the shared space.
The more advanced scribe, S4 described their job as filtering information for clinicians based on relevancy and importance.
Less advanced scribes perceived their role as recording rather than synthesizing or finding information.

Scribes integrated cards into documentation and retrieval processes for multiple concept types. For example, scribes mentioned using lab cards to compare a patient's current value to their baseline, clicking on a past procedure chip to pinpoint its date from surfaced notes, and leveraging condition cards to determine the extent and severity of a patient's existing condition. This indicates such cards act as information scent to guide scribes to important content. S4 described that they would use cards to dig into particularly relevant medical history; for example, if a patient with chest pain had past cardiac disease, the scribe would utilize the card to review "their previous work-up, notes from cardiology, and any prior surgeries/procedures." S3 noted particular utility in associated medications that were surfaced on condition cards; it prompted them to document, and the concept-oriented view also served an educational purpose of teaching them what was relevant, potentially aiding future synthesis. 

While multiple scribes noted it was less useful for medications and symptoms, condition cards aided scribes in understanding the extent and trajectory of a patient's past condition. S4 noted that they "use it when... interested in more information about a patient's medical history, especially in a complex patient or a patient that is unable to provide a history due to acuity or altered mental status."

During the deployment \mytitle proactively displayed 3614 cards to scribes as they typed, with a range of frequencies. For example, 53 hypertension cards were surfaced after typing "htn", while 144 terms were displayed only once each; these rarer terms included "spine fracture" and "lumbar spinal stenosis". Some short phrases that overlap with common language e.g. \code{as} as \code{atrial stenosis} were mistakenly tagged as clinical terms, but future iterations of \mytitle can recognize these and omit them.  

Experienced scribes tended to familiarize themselves with a patient's medical history prior to writing a note, and thus used the search bar to display cards prior to note-writing rather than triggering them automatically during the course of documentation. Future iterations of \mytitle may want to support this workflow, since the existing information capture is focused on displaying information as the physician types rather than before the physician starts the note, which misses an opportunity to provide information scent before documentation begins. Less experienced scribes were more likely to click on chips within the note to see relevant cards. All scribes used chips to manually view cards in the sidebar, and all but one (S4) used chips as the primary tool for bringing up cards in the sidebar. The usage data reveals that users are willing to adopt a wide variety of techniques for accessing documentation, but appear to have significant preferences for one technique or the other.

\section{Discussion}
\mytitle explores several interaction paradigms by enabling live automatic recognition of clinical terms within a medical note and displaying patient medical history in concept-oriented cards.
Our iterative design process for \mytitle underscored the need for EHR systems to embrace and augment, rather than replace existing clinical workflows.
Our features were well received when they synergized with existing documentation practices.
Implementing changes in a clinical environment is challenging, and clinicians and scribes are more receptive to such changes when presented with tools that are familiar and intuitive.

In future iterations of \mytitle, we hope to expand on the possibilities enabled by fine grained linking of chips, in both the note and card interfaces, to standard medical ontologies.
\mytitle can leverage existing health knowledge graphs or outside resources that clinicians use, aiding their decision-making during documentation. 
Normalization to a standard ontology also allows notes to be translated to different audiences; medical acronyms can be automatically unravelled to layman's terms if a patient wants to understand their note.
Clinicians with specific language preferences can also personalize note templates and autocomplete functionality with the vocabulary choices that they prefer. 

Our observations from user interviews and interaction data have additionally presented new avenues for future work that could push forward these interaction paradigms. Clinicians often chunk information together. When a clinician wants to view a \emph{Hemoglobin} lab, they are likely to search for \emph{CBC} (Complete Blood Count), a set of laboratory tests, since \emph{Hemoglobin} is recorded as part of a \emph{CBC lab group}. \mytitle could support such lab groups by adopting a wider clinical ontology, or even allowing clinicians to merge or combine cards within the user interface, dragging and dropping multiple lab cards together to create higher level lab groups. These modifications do not have to be limited to labs. Clinicians could, for example, group a glucose card with a diabetes card because the medication is directly treating th condition.

Providing clinicians with the ability to mold their information displays could not only help physicians synthesize medical records, but also create new possibilities for crowdsourcing rich labeled datasets of clinical relationships. Clinician-curated content would also be a potential solution for how to scale from a handful of manually curated cards to thousands of cards, and even create cards that serve different roles for different types of users (less granular for generalists like emergency physicians or primary care physicians, more granular for specialists like oncologists or immunologists). 

In the opposite direction, clinicians sometimes want to refer to a specific value or event when recording information. When the clinician writes "patient has high glucose" it would be helpful if the system identified exactly which glucose lab was high, autocompleting not just glucose as a clinical term, but a specific measurement. By allowing clinicians to refer to granular as well as chunked information, we can get closer to the ideal of presenting information to the clinician in a way that mirrors their clinical thought process.

Cards could be improved by offering further custom views of the patient medical record, and providing context-dependent defaults. For example, the existing lab cards in \mytitle display a table of result by default, but can display a line chart, and box and whisker plot as well. For some labs the most recent value is the only value that matters, and a table is appropriate, but for other labs, the trend over time is what matters, and a line chart would be more useful. A clinician's mental model of the patient becomes more refined over a patient's visit, resulting in different information needs as the visit progresses. An area of future work would be to investigate how to support this change. Contextual display of information in cards is challenging, but would continue to shift some of the cognitive burden of synthesizing the patient medical record onto the EHR.

We noted several occasions where usage differed based on clinical experience. However, even advanced clinicians can be novice users of the tool --- experience level with \mytitle is thus another dimension of the overall user experience. Ideally, a user interface would be intuitive for the novice user and provide support to help them grow into advanced users for the tool. Future work could expand on the logging we have implemented here to see how a user's usage of the tool changes over time, and what strategies we could employ to improve adoption of more advanced features.

There is a practical burden in scaling cards.
However, because we use standard ontologies, we can leverage ongoing efforts to open-source physician-curated \cite{semanikImpactProblemorientedView2021} and machine learned \cite{mullenbachKnowledgeBaseCompletion2020} concept maps.
Users may benefit from the ability to manually author and customize default text templates and card contents but we hypothesize that a relatively small set of cards could cover most referenced terms.
\textit{Semanik et al.} estimate that 150-200 concept maps could be used to cover the most commonly encountered conditions for a range of clinical specialties \cite{semanikImpactProblemorientedView2021}.

There is some risk that the automation provided by \mytitle could lead to errors. For example a post recognition may be incorrect, modifiers may be skipped or added unnecessarily, and auto populated text could contain errors.
Auto populated text requires manual verification, but this is an existing step in physician workflows, since the current system provides naive boilerplate text for modification.
The risk of incorrect tags is lower than comparable clinical recognition systems \cite{HealthcareNaturalLanguage2021,savovaMayoClinicalText2010} because users are able to disambiguate terms and have information scent about the recognized terms.

There is additional risk that adopting \mytitle could impede usability.
To that end we have designed \mytitle' features to be opt-in, leaving existing workflow unimpeded.
Lastly we must consider if using \mytitle could impact the responsible practice of medicine.
The literature considers the risks of passive clinical decision support (CDS) like our concept-oriented cards to be minimal compared to active CDS \cite{semanikImpactProblemorientedView2021}.
While cards provide synthesized evidence, the responsibility is on the clinician to explore as needed, and data on cards links to the original source, e.g. note snippets expand to the full note.

\section{Conclusion}
\mytitle captures structured clinical terms embedded within a free-text narrative and then links these terms to a concept-oriented, dynamic display of patient information that appears alongside a medical note. Thus, \mytitle provides clinicians with a unified interface for writing a clinical note and exploring and navigating a patient's medical record. By integrating documentation and patient information \mytitle lowers the cognitive burden of synthesizing the medical record, and demonstrates the possibilities of an EHR documentation system that can better serve clinicians. 

We capture these clinical concepts via \emph{autocomplete} and \emph{post recognition}, and map them to standardized ontologies. This allows us to connect captured concepts with other medical databases and translate clinical terms for a variety of audiences. We provide patient information in a preview pane next to the note as the clinician types, proactively displaying contextual information when needed. A persistent sidebar of cards helps multiple clinicians develop a shared understanding of the patient and highlights important information. We prove the feasibility of our approach in a months-long deployment in an active ED, and demonstrate in our evaluation that clinicians are receptive to this approach. 

Ultimately, we believe that \mytitle has the potential to make clinical documentation truly work for clinicians by creating a live document that supports customized information retrieval, note-taking, and collaboration while simultaneously improving the final note that is shared with downstream doctors and patients.

\begin{acks}
We would like to thank Rachel Feldman and the other scribe users as well as Nicholas Kurtzman and other clinicians who provided valuable feedback. This project would not have been possible without the help and patience of the scribes and clinicians who were willing to put up with many bugs in our early deployment. This work was supported by an award from the MIT Abdul Latif Jameel Clinic for Machine Learning in Health (J-Clinic).
\end{acks}

\bibliographystyle{ACM-Reference-Format}
\typeout{}
\bibliography{references}










\end{document}